\documentclass[floatfix,aps,reprint,prd,nofootinbib,amsfonts,amsmath,amssymb]{revtex4-2}
\usepackage{geometry}        
\usepackage{graphicx}
\usepackage{amssymb}
\usepackage{amsmath}
\usepackage{epstopdf}
\usepackage{xcolor}
\usepackage{subfigure}
\usepackage{hyperref}

\newcommand{\beq}{\begin{equation}}
\newcommand{\eeq}{\end{equation}}

\begin{document}

\title{Where are all the dark galaxies? \\Predicting galaxy/halo locations from their bright neighbors}

\author{Alice Y. Chen}
\email{ay7chen@uwaterloo.ca}
 \author{Niayesh Afshordi}
 \email{nafshordi@pitp.ca}
\affiliation{Department of Physics and Astronomy, University of Waterloo, 200 University Ave W, N2L 3G1, Waterloo, Canada}
\affiliation{Waterloo Centre for Astrophysics, University of Waterloo, Waterloo, ON, N2L 3G1, Canada}
\affiliation{Perimeter Institute For Theoretical Physics, 31 Caroline St N, Waterloo, Canada}

\begin{abstract}
Astronomical objects in our universe that are too faint to be directly detectable exist and are important - an obvious example being dark matter.  The same can also apply to very faint baryonic objects, such as low luminosity dwarf galaxies and gravitationally compact objects (e.g., rogue planets, white dwarfs, neutron stars, black holes, dark sirens).  While they are very difficult to observe directly, they have locations that are highly important when studying astrophysical phenomena.  Here, we use a machine learning algorithm known as symbolic regression to model the probability of a dark object's existence as a function of their separation distances to their closest two ``bright" (directly observable) neighbors, and the distances of these bright objects to each other.  An advantage of this algorithm is that it is interpretable by humans and can be used to make reproducible predictions.  Galaxies with masses above $10^9 M_{\odot}$ and halos above $10^{12} M_{\odot} $ are the objects that we separate into ``bright" and ``dark" to be used in our analysis.  We find that it is possible to predict the density of dark objects using an analytic expression that depends on their distances to their closest bright neighbors in Illustris-TNG galaxy formation simulations, which is significantly better than the (linear) scale-dependent biasing prediction for $k \sim 1.0~ h$Mpc$^{-1}$ (and potentially beyond, if allowed by the resolution).  This could potentially open the avenue for finding dark objects based on their vicinity to directly observable bright sources and make future surveys more targeted and efficient.
\end{abstract}
\maketitle

\section{Introduction}
\label{sec:intro}

Localizing faint discrete objects is important for studying many astrophysical phenomena.  A well-known example is using the location of ``dark sirens" -- binary compact object mergers that produce gravitational waves without an electromagnetic counterpart -- to measure the Hubble constant \cite{NANOGrav:2023gor}.  These sirens often reside in galaxies, which in turn are hypothesized to reside in larger dark matter structures known as halos.  Although millions of galaxies have been observed using their optical luminosity, halos are composed mainly of dark matter, which emits little to no light.  Consequently, localizing halos, and even faint galaxies residing within halos, becomes a difficult challenge.

What is known is that halos and galaxies form by gravitational collapse, clustering, and hierarchical assembly \cite{Cooray:2002dia}.  Therefore, it is reasonable to expect a pattern in their distributions - reflected in their separation distances - since close-by neighboring objects would gravitationally influence each other and assemble from small to large.  There have been many previous works exploring halo models and large scale matter distribution based around the theory of matter collapsing into quasi-static virialized objects (e.g. see \cite{Cooray:2002dia, Asgari:2023mej} for reviews).  Most of this previous work was aimed at exploring properties such as halo bias \cite{2010ApJ...724..878T}, mass function (e.g., \cite{Tinker:2008ff}), halo concentration (e.g., \cite{Okoli:2017uts}), and N-point correlation functions/power spectra (e.g., \cite{Cooray:2002dia, Asgari:2023mej}).  More recent work on cosmological large-scale structure involves using machine learning to recover initial conditions from simulation data based on halo clustering (e.g., \cite{Legin:2023jxc}), interpreting halo profiles once structure has formed \cite{Lucie-Smith:2022uvv, Lucie-Smith:2023kue}, and exploring galaxy and stellar population in halos \cite{Wu:2024jyu}.  However, looking at the halo distribution solely as a function of separation distance to a halo's $k$ nearest neighbors has not yet been explored in detail.  In this study, we provide a model of the locations of `dark' halos and galaxies based on their separation distances to their nearest two `bright' neighbors and the neighbors' distances to each other; this is similar to a conditional 3-point correlation function (the condition being the two bright objects being the closest neighbors of a dark object).  Considering the gravitational effects that halos and galaxies exert on one another and their interconnected history, it stands to reason that a mathematical function should exist to determine the likelihood of detecting structures in close proximity based on their separation distance. Our objective here is to develop a model for this likelihood.

To do this, we use the PySR symbolic regression package \cite{Cranmer:2020wew,2023arXiv230501582C} as our main modelling tool.  Although standard parametric fitting algorithms require the assertion of an ad hoc parametric form \cite{harris2020array, 2020SciPy-NMeth}, symbolic regression can systematically find a suitable parametric form \cite{koza1992programming, Schmidt2010, Cranmer:2020wew, 2023arXiv230501582C}, and thus can accurately model data of higher complexity.  This is especially useful when there is not much prior information, as is the case for the distributions of dark haloes.  Furthermore, the parametric form of symbolic regression outputs is easily interpretable for humans, whereas trained neural networks, though accurate, appear as a black-box.

Here, we apply the PySR symbolic regression algorithm to results from the Illustris-TNG collaboration \cite{Nelson:2018uso, Vogelsberger:2014dza} to model galaxy and halo distributions on the scales of nearest neighbors.  Illustris-TNG is an N-body cosmological simulation that uses a moving particle-mesh grid to model the gravity and hydrodynamics of particles.  The data we use here are from the Illustris-2 - with a volume of $(75~ {\rm Mpc/h})^3$, $2\times 910^3$ dark matter and gas particles  - and TNG-100 - with the same volume but 8 times more particles.  The simulation results have redshifts of 0 to 128, use Planck 2015 cosmology \cite{Planck:2015fie}, and contain both halos and subhalos, which are equivalent to satellite galaxies within the halos.

This paper is then sectioned as follows - the methods used and our application of Poisson probability are discussed in Section \ref{sec:background}, the  results are shown in Section \ref{sec:results}, and a summary of the results and potential future work is outlined in Section \ref{sec:conclusion}.

\section{Statistical Methods and Modeling}
\label{sec:background}

\subsection{Distance Calculations and Correlation}
\begin{figure}
    \centering \includegraphics[width=1.0\linewidth]{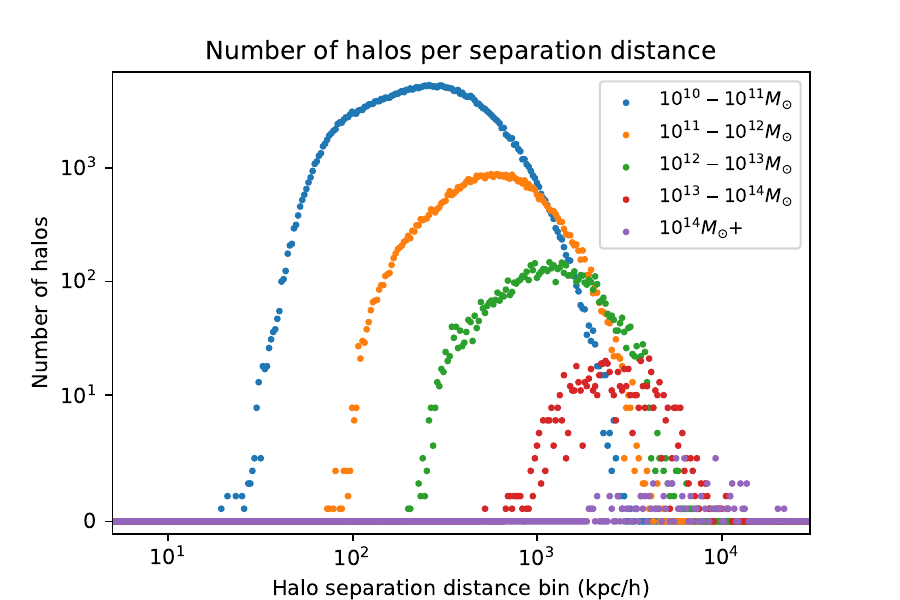}
    \caption{\footnotesize Separation distances (co-moving) of halos from their closest neighbors, split by mass bins.  Note that not all the halos in this figure are used in our analysis - we only include halos more massive than $10^{12}M_{\odot}$, as smaller haloes may be significantly impacted by numerical resolution of the simulation.}
    \label{fig:sep_1}
\end{figure}

Most information we have about dark matter halo clustering and distribution have come from large scale cosmological simulations \cite{Angulo:2021kes}, with GADGET \cite{Springel:2000yr}, Millenium \cite{2005Natur.435..629S}, and Illustris-TNG, \cite{Vogelsberger:2014dza} being well-known examples.  Using the latter, we see that the distance to the closest neighbor of a halo roughly follows a log-normal probability distribution, with the average separation distance increasing for more massive\footnote{with mass defined as the total mass within a radius with $200$ times the critical density of the universe. } -- and thus sparse -- halos (see Figure \ref{fig:sep_1}).  This mass-dependent trend agrees with our current understanding of halos: more massive halos collapse from larger regions and thus are less densely concentrated \cite{1996ApJ...462..563N}.

Although previous work has often quantified halo clustering in terms of bias and mass functions \cite{Cooray:2002dia, Tinker:2008ff, 2010ApJ...724..878T, Asgari:2023mej}, these descriptions are only applicable on very large scales and difficult to observe empirically at small separations.  Given that the universe, and simulations, are translationally invariant, we will instead study the probability of finding a galaxy or halo in the vicinity of another neighboring object in the simulation.  To do this, we bin the lower-mass and less luminous ``dark'' objects based on the smallest two separation distances from their ``bright'' neighbors.

With these dark and bright object samples, we then try to find an analytic function that quantifies the density of dark objects based on their separation distances to their closest bright objects, as well as the distance of the bright objects to each other.  This is similar to the method of $k$-nearest neighbor ($k$NN) statistics \cite{Banerjee:2020umh}, except that here we constrain to $k=2$ for simplicity and consider the distances to dark objects, rather than random Poisson points. 

\begin{figure*}
    \centering \includegraphics[width=0.98\linewidth]{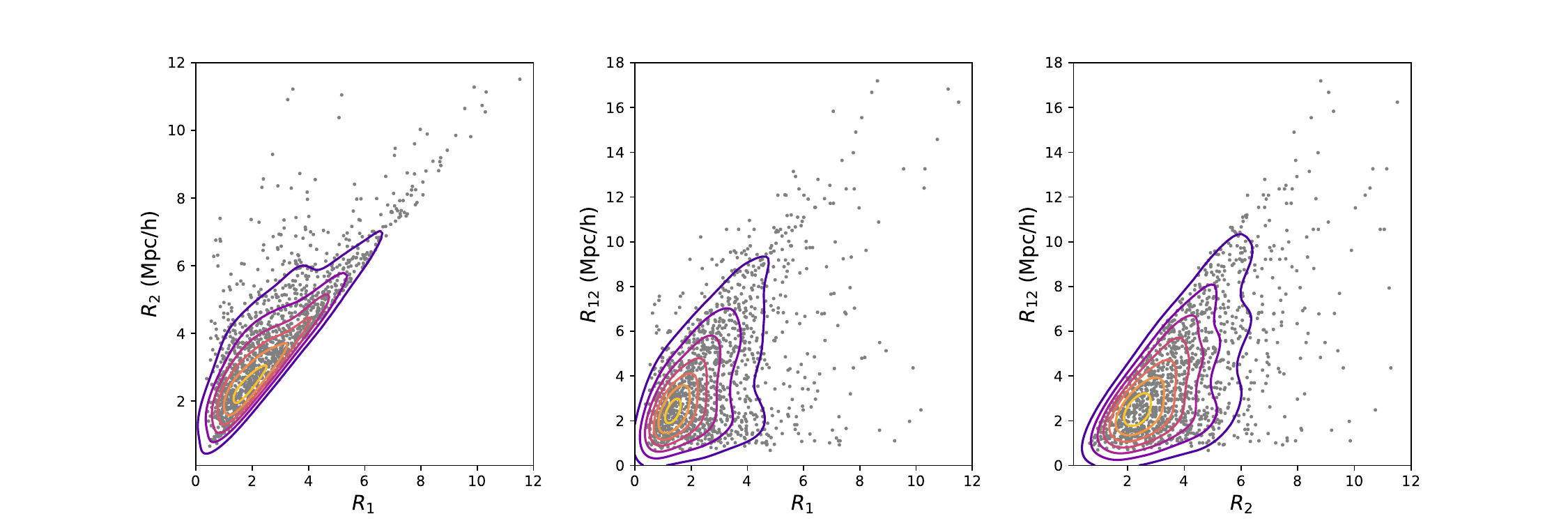}    \includegraphics[width=0.98\linewidth]{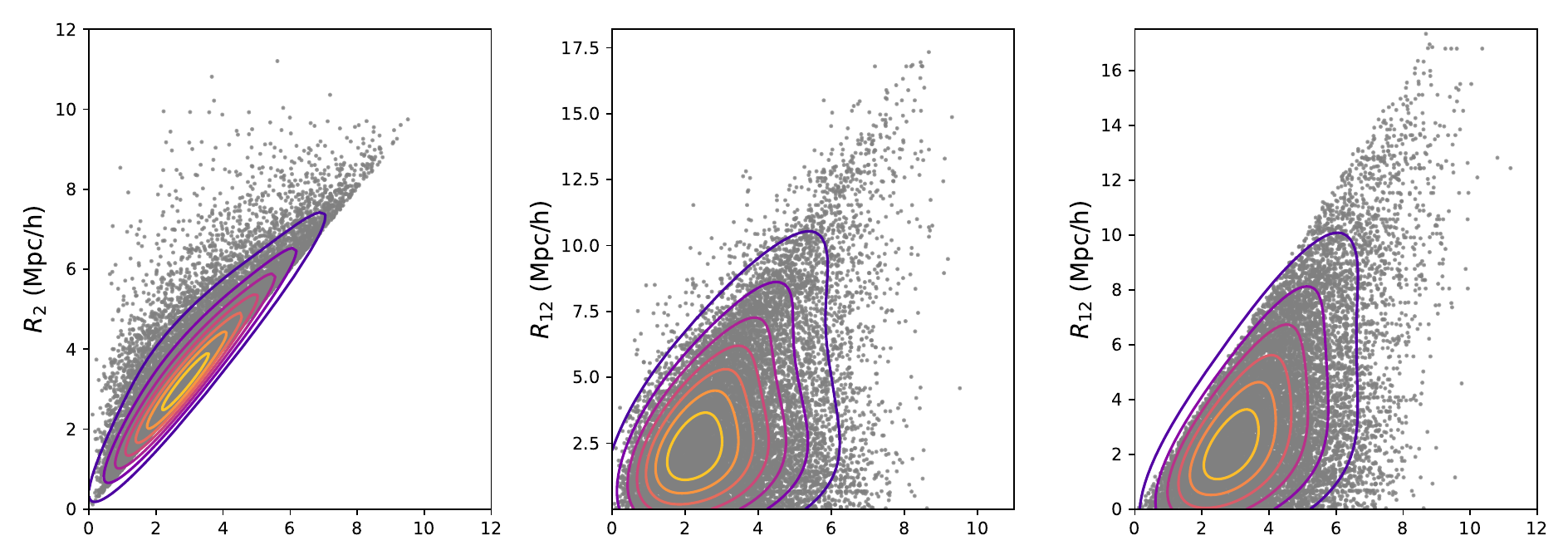}  \includegraphics[width=1.0\linewidth]{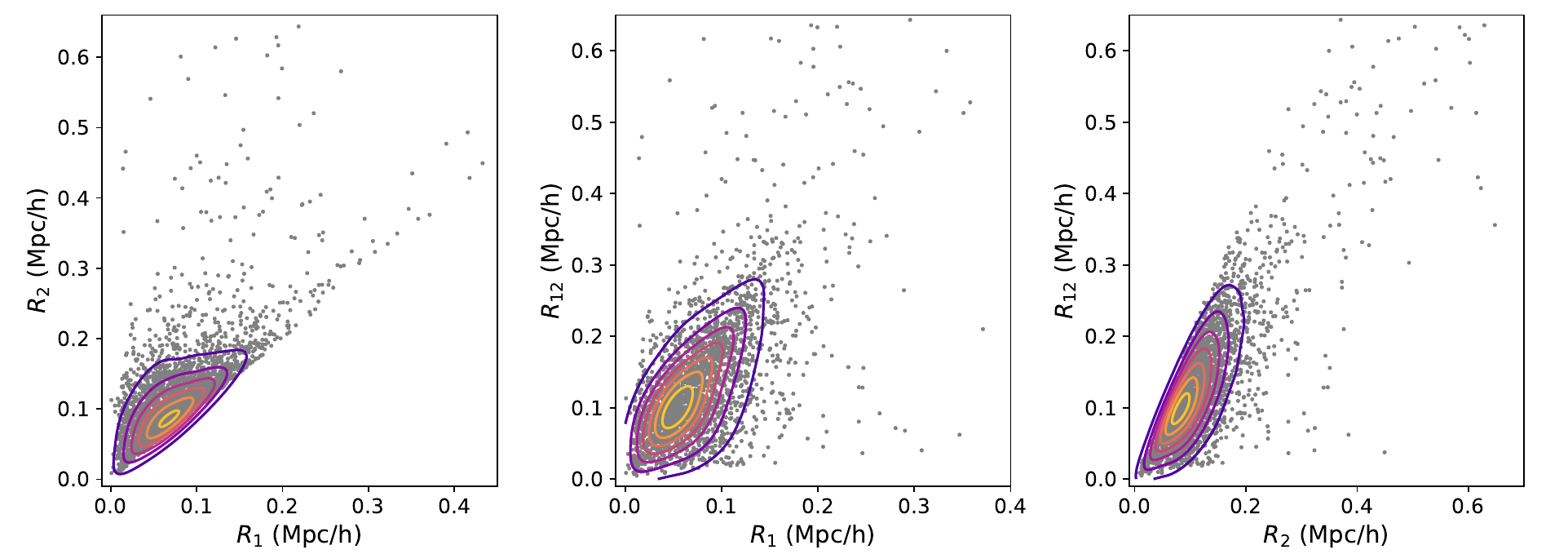}
    \caption{\footnotesize 
    The distribution of distances of ``dark objects'' to to their closest bright neighbor, $r_1$, their second closest bright neighbor $r_2$, and the separation of these two bright neighbors $r_{12}$. The ``dark objects'' are taken to be low-mass galaxies in the \textit{Middle} panel, while they are dark matter haloes in the  \textit{Top} and \textit{Bottom} panels. Meanwhile, the ``bright neighbors'' are assumed to be massive galaxies in \textit{Middle} and \textit{Bottom} panels, while they are dark matter haloes in the \textit{Top}. The isodensity contours enclose, $\sim$15\% to $\sim$85\% of the points, from smallest to largest, respectively. The sharp edges in the distribution are imposed by triangle inequalities. 
    }
    \label{fig:halo_contours}
\end{figure*}

To obtain a probability we take the total number of halos in the simulation and divide them based on their distance to their closest neighbors while imposing a mass cut-off at $10^{12} M_{\odot}$ (see distance contours in Figure \ref{fig:halo_contours} and probability densities in Figures \ref{fig:sep_ml}, \ref{fig:sep_sub}, \ref{fig:sep_halo_galaxy}).  We then divide the number of halos in each bin by the total number of halos available to obtain the probability of finding a halo in each distance bin.  The distance between halos is found using the Euclidean distance formula 
\beq 
r=\sqrt{(x_2 - x_1)^2 + (y_2 - y_1)^2 + (z_2 - z_1)^2}
\eeq
where $x_1, y_1, z_1$ and $x_2, y_2, z_2$ are the positions of a halo and its closest neighbour, respectively.  The distance calculation procedure takes into account the simulation's periodic boundary conditions by including halos on the other side of the simulation box when calculating separation distances.  

While we know the exact masses of halos from simulations, in real observations we don't have an accurate mass measurement. Therefore, we separate ``dark'' and ``bright'' halos by random sampling of halos $\geq 10^{12}M_{\odot}$. As for galaxies, we define the bright ones as having $M\geq 10^{10} M_{\odot}$, and dark galaxies as $2\times 10^9 M_{\odot} \leq M \leq 10^{10} M_{\odot}$, where $M$ is the total mass of the galaxy/subhalo.  

We then sort the range of separation distances between dark objects and their two brightest neighbors into bins, and model the probability density of a dark object being in each distance bin using PySR. 

\subsection{Poisson Distribution}

It can be seen from Figure \ref{fig:sep_1} that the number of halos in each separation bin roughly resembles a skewed log normal.  Let us start with the null hypothesis that halos are randomly selected from a uniform 3D distribution; however, we shall later observe that gravitational effects break this assumption, as illustrated in Figures \ref{fig:sep_ml}-\ref{fig:sep_halo_galaxy}.

The general Poisson probability distribution takes the form
\beq 
P(x) = \frac{\lambda^x}{x!}e^{-\lambda},
\label{eq:Poisson}
\eeq 
where $x$ represents the number of halos in a given volume and $\lambda = \langle x \rangle = {\bar n}_b \times~ {\rm volume}$, where ${\bar n}_b$ is the mean number density of bright haloes.

Let us define the probability density:
\beq 
p(r_1, r_2, r_{12}) \equiv \frac{dP(r_1, r_2, r_{12})}{dr_1 dr_2 dr_{12}}, 
\eeq
where $r_1$ is the distance to closest halo/galaxy, $r_2$ is the distance to second closest halo/galaxy, and $r_{12}$ is the distance between the two closest halos/galaxies. For a three dimensional sprinkling of points at uniform density, we can combine Poisson probability (\ref{eq:Poisson}) with Bayes's theorem to find
\begin{eqnarray}
p(r_1, r_2, r_{12}) = p(r_{12}|r_1,r_2)p(r_2|r_{1})p(r_{1}) \nonumber\\ =\frac{r_1 r_2 r_{12} (4\pi \Bar{n}_b)^2 e^{-\frac{4}{3}\pi r_2^3 \Bar{n}_b}}{2} \Theta(r_1, r_2, r_{12}),
\label{eqn::prob}
\end{eqnarray}
where 
\begin{eqnarray}    
   &&\Theta(r_{12}, r_1, r_2) = \nonumber\\&&\begin{cases}
    1, & \text{if } (r_2 - r_1) \leq r_{12} \leq (r_1 + r_2) \\
    0, & \text{otherwise},
\label{eqn::theta}
\end{cases}\nonumber\\
\end{eqnarray}
enforces triangle inequality. 

To plot Eq. \ref{eqn::prob} with respect to each variable $r_1, r_2, r_{12}$, we can project (i.e. integrate) the probability over the two other separation distances.  The integral over $r_{12}$ and $r_2$ while keeping $r_1$ as the variable is
\beq 
\frac{dP}{dr_1} = 4\pi R^2_1 \Bar{n}_b e^{-\frac{4}{3}\pi R^3_1 \Bar{n}_b},
\eeq 
while the integral over $r_{12}$ and $r_1$, keeping only $r_2$ is
\beq 
\frac{dP}{dr_2} = \frac{16 \pi}{3} r^5_2 \Bar{n}_b^2 e^{-\frac{4}{3}\pi r^3_2 \Bar{n}_b},
\eeq 
and the integral over $r_2$ and $r_1$, keeping $r_{12}$ is 

\begin{eqnarray}    
&&\frac{dP}{dr_{12}} = \nonumber\\ &&2\pi \Bar{n}_br^2_{12} e^{-\frac{\pi \Bar{n_b} r^3_{12}}{6}}
 - \frac{r^5_{12} \Bar{n}_b^2 \pi^2}{3} E_{1}\left(\frac{\pi \Bar{n}_b r^3_{12}}{6}\right),\nonumber\\
\end{eqnarray} 
where $E_1$ is the exponential integral of rank 1.

\begin{figure*}
    \centering    \includegraphics[width=1.0\linewidth]{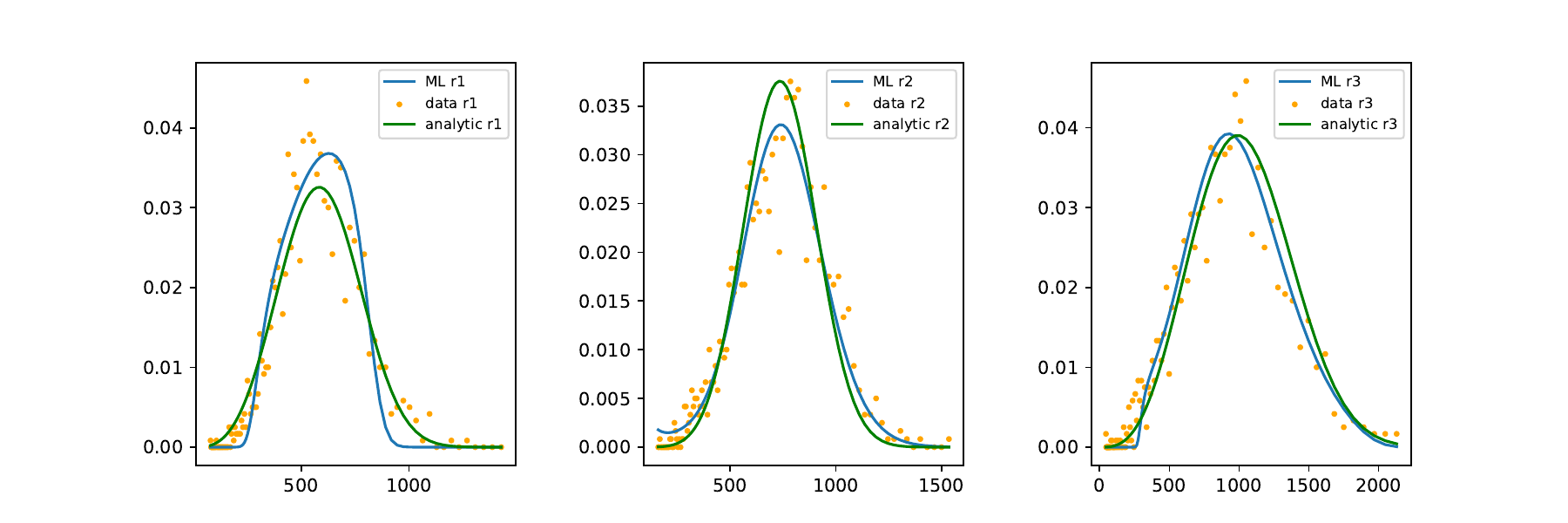}
    \caption{\footnotesize Random sampling in general, to compare the ML algorithm (symbolic regression) against the analytic Poisson expressions above and see how well they perform.  If average number density $\Bar{n} \sim 10^{-12}$ and the number of hypothetical dark halos are on the same scale as the number of bright halos, then the analytic expressions match the data reasonably well.  However, the actual distribution of objects in the universe is not random, and it will be seen later that using only the Poisson distribution fails to model the data accurately.}
    \label{fig:poisson_test}
\end{figure*}

This is the starting assumption used to model the distribution of neighbor distances. While it is good for random data (see Figure \ref{fig:poisson_test}), it can be seen below (see Figures \ref{fig:sep_ml}, \ref{fig:sep_sub}, \ref{fig:sep_halo_galaxy}) that a random Poisson distribution by itself is not enough to predict simulated astronomical object location probabilities, due to the effects of gravitational clustering.  This shows the need to develop a more predictive model, which motivates the rest of this work.

\subsection{Poisson Error Analysis and Maximizing Likelihood}
Symbolic regression returns many potential mathematical functions that can be used to model the probability density of ``dark'' objects in the vicinity their ``bright'' counterparts (see Table \ref{tab:all_eqns}).  PySR aims to find the best fitting function by minimizing its error.  In machine learning, the error is calculated using a loss function, the most common of which is the mean square error (MSE) function
\beq 
{\rm MSE} = \frac{1}{n}\sum^n_{i=1} \Big( y_{\rm prediction} - y_{\rm data} \Big)^2.
\label{eqn::mse}
\eeq 
This loss function works best for data that follow a Gaussian error distribution, but for discrete objects such as galaxies/clusters, Poisson probability is more appropriate.  Therefore, we define our own loss function, where we maximize the probability of finding an object (maximizing Eq. \ref{eq:Poisson}).  This is equivalent to minimizing its inverse, which results in a loss function of
\beq 
L = \prod_{i=1}^n\frac{x_i!}{\lambda_i^{x_i}}e^{\lambda_i}.
\label{eqn::poisson_loss}
\eeq 
 In our analysis, we compute $\lambda_i$, {\it for every voxel (of volume $\Delta V$) in the simulation}, as a function of $r_1$, $r_2$, and $r_{12}$ for {\it the centre} of that voxel, then find $\lambda(r_1, r_2, r_{12}) = p(r_1, r_2, r_{12}) \Delta V$ such that it minimizes the loss function (\ref{eqn::poisson_loss}) for a given simulated number of dark galaxies/haloes per voxel $\{x_i\}$. The resulting $p(r_1, r_2, r_{12})$ should predict our dark object density with minimal errors.  
To take the product into a sum, we can use its logarithm, and define
\begin{widetext}
\beq 
\tilde{L} \equiv \sum_{i=1}^n \Bigg[ \lambda(r^i_1,r^i_2,r^i_{12}) - x_i\bigg(1 - \ln\left| \frac{\lambda(r^i_1,r^i_2,r^i_{12})}{x_i}\right| \bigg) \Bigg].
\eeq 
\end{widetext}
While $\tilde{L} = \ln(L)$ up to a constant that does not depend on $\lambda$, it is positive semi-definite and vanishes only if the data exactly match the model, $\lambda(r^i_1,r^i_2,r^i_{12})=x_i$, just like MSE (Eq. \ref{eqn::mse}).  Although the number of galaxies/haloes per voxel is $x_i \geq 0$, the PySR algorithm fit for $\lambda$ is not necessarily $\geq 0$ for every voxel.  To ensure that this does not become a problem for the $\ln$ function, we can make a change of the variable $\lambda \rightarrow e^{\phi}$, where we fit for $\phi$ instead.  The loss function then becomes
\begin{widetext}
\beq 
\tilde{L} \to \sum_{i=1}^n \Bigg[ e^{\phi(r^i_1,r^i_2,r^i_{12})} - x_i \bigg( 1 - \ln\left| \frac{e^{\phi(r^i_1,r^i_2,r^i_{12})}}{x_i+\epsilon}\right|\bigg) \Bigg],
\label{eqn::L_exp}
\eeq 
\end{widetext}
where $\epsilon$ is a softening length to avoid an undefined $0\ln|0|$ in the code (we set $\epsilon$ to be $10^{-10}$), although it has no effect on $\tilde{L}$. 

The list of equations (also known as the ``Pareto front'') that PySR has given us is shown in Table \ref{tab:all_eqns} in the Appendix.  Not all separation distances, $r_1, r_2, r_{12}$ are used in the equations, and sometimes the combination is not very physical, since there is dependence on the second closest neighbor ($x_1 = r_2$), but not the first closest neighbor ($x_0 = r_1$).  From how gravity works, this does not sound plausible, as the closer an object is to another, the stronger its gravitational effect.  Therefore, we exclude the equations that do not have a dependence on $x_0$, and focus only on those that do in our error analysis below.

\begin{figure}
   \centering \includegraphics[width=1.0\linewidth]{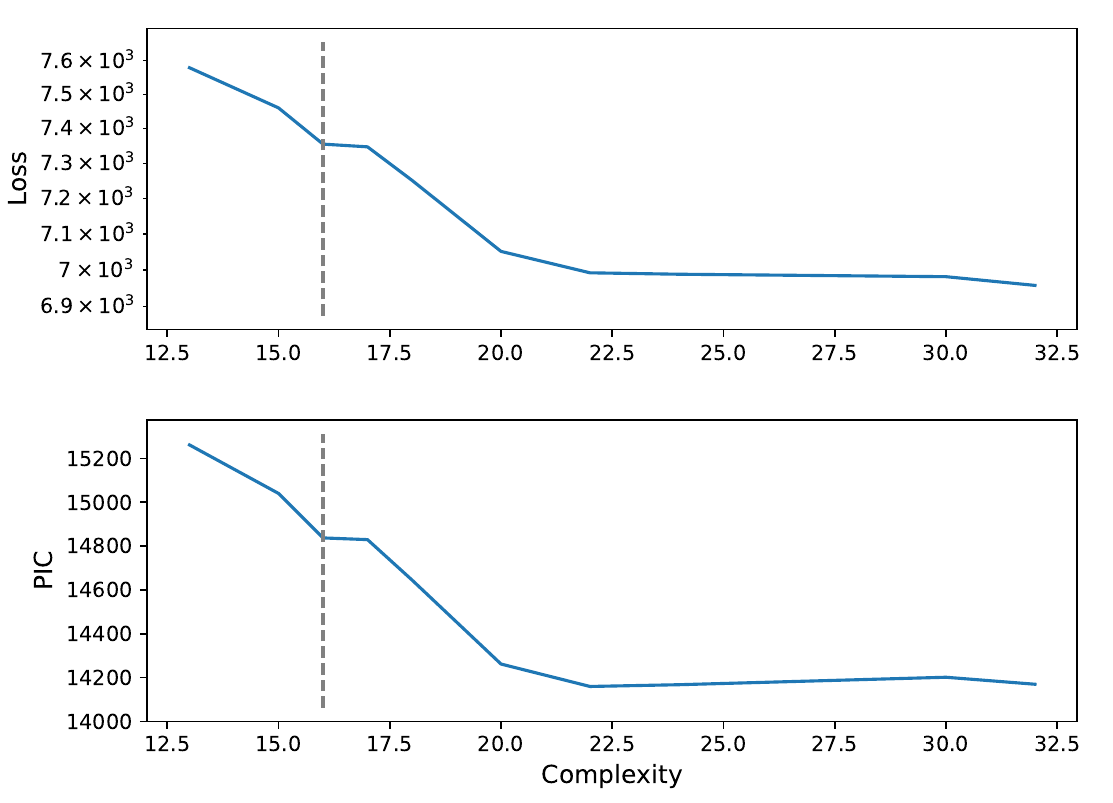}
   \caption{\footnotesize  The loss, Eq. \ref{eqn::L_exp} (top) and Poisson Information Criterion, PIC (Eq. \ref{eqn::PIC}; bottom) for each PySR fit function, as a function of its complexity.  The local minimum with a large decrease in loss (high slope) is where we select the preliminary fit for the galaxy probability distribution (gray vertical line).  This occurs where the PIC is also decreasing the fastest with regards to complexity.  Since these two test coincide at complexity=16, the unique function with that complexity is the one we use for our probability model in Eq. \ref{ml_func}.}
\label{fig:halo_loss}
\end{figure}

Using the loss function from Eq. \ref{eqn::L_exp}, and equations 4, 8, and 14 from Table \ref{tab:all_eqns}, we then pick the function from PySR that has the largest decrease in loss with the increase in complexity (i.e., the largest negative slope).  This is for mathematical simplicity, since even with complexity constraints, symbolic regression will return many equations that could fit the data.  This is shown in Figure \ref{fig:halo_loss}, where we pick the function with complexity and loss on the gray vertical line.  

Further tests of how good of a fit each function is are the Bayesian Information Criterion (BIC), Akaike Information Criterion (AIC) \cite{2013arXiv1307.5928G}, and the minimum message length (MML) \cite{WONG20181}.  Let us focus on BIC, defined as
\beq 
{\rm BIC} \equiv -2\ln[P(\lambda|L)] + k\ln(n),
\label{eqn::BIC}
\eeq 
since it is the closest approximation to Bayesian evidence, with a concrete statistical interpretation.  Here, $k$ is the number of parameters in the probability model, and $n$ is the number of data points.  

However, this definition of BIC is only well-suited for Gaussian errors on data and parameter posteriors.  Since we are starting with a Poisson distribution, we adapt Eq. \ref{eqn::BIC} to be what we now call a ``Poisson Information Criterion" (PIC), which takes the form
\beq 
{\rm PIC} = -2\ln[P(\lambda|L)] + k \ln(n_{\rm object}),
\label{eqn::PIC}
\eeq
where $n_{\rm object}$ is the total number of objects in the simulation (bright + dark) and we take $k$ as the complexity, defined as the number of nodes of each expression tree \footnote{This is the the sum of the number of times each operator appears + the number of times each constant in front of a variable appears \cite{Cranmer:2020wew}}, used by PySR.  The results of this are shown in the bottom panel of Figure \ref{fig:halo_loss} along with the results of the loss function. 

\section{Results}
\label{sec:results}

\begin{figure}
    \centering    \includegraphics[width=0.95\linewidth]{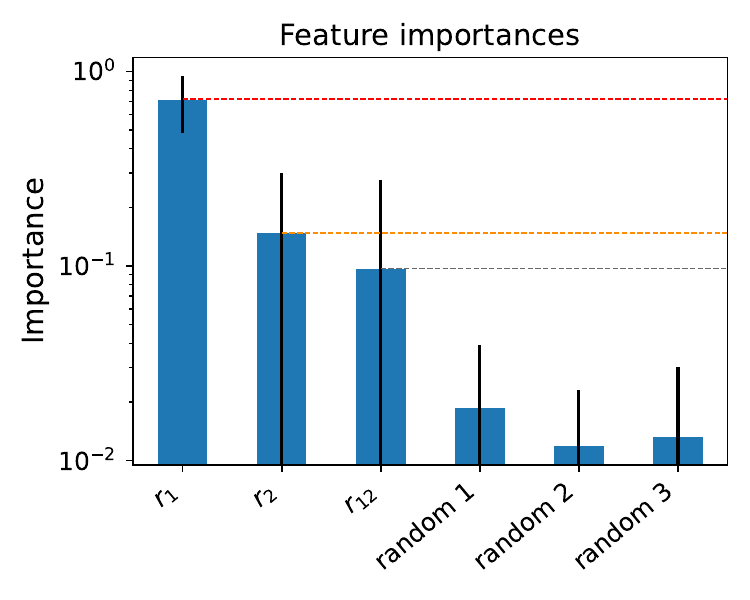}
    \caption{\footnotesize Comparing the importances of separation distance vs. random numbers on halo distribution.  Here we see that the separation distances have significant importance, as expected due to gravitational clustering.}
    \label{fig:rand_forest}
\end{figure}

To calculate the probability of finding a halo or galaxy object in a certain volume element, we model simulation data points by using symbolic regression and the Poisson error function in Eq. \ref{eqn::L_exp} as the loss function to be minimized.  We find that it is possible to find a fitting function that models the data well using this method (see top graph of Figure \ref{fig:halo_loss}).  However, since this is likely true for any set of variables in general, one way to test the significance of separation distances is to compare them to the effect of random numbers on halo/galaxy distributions using a random forest classification, which can quantify the significance of each variable on another \cite{8122151}.  The results of this are shown in Figure \ref{fig:rand_forest}, where we see that the separation distances $r_1, r_2, \text{and } r_{12}$ are indeed more significant than the random numbers which should have little to no effect on halo/galaxy distribution.  This agrees with the expected result since halos and galaxies tend to cluster around each other, and shows that the probability of finding a halo or galaxy does have some dependence on its closest neighboring distances to other halos/galaxies. We use the subhalo galaxies as our training set and host halos as our testing set.

The probability density function $p(\Vec{r})$, where $\Vec{r}=(\Tilde{r}_1, \Tilde{r}_2, \Tilde{r}_{12})$, that was found for the number of objects as a function of separation distances is 
\beq
p(\Tilde{r}_1, \Tilde{r}_2, \Tilde{r}_{12}) = e^{\frac{1}{a\Tilde{r}_2 + b}} \Big( \frac{c}{2\Tilde{r}_2 + \Tilde{r}_{12}} \Big)^{d\Tilde{r}_1 + e}
\label{ml_func}
\eeq 
where $a, b$, $c$, $d$, and $e$ are nuisance parameters used for fitting.  $\Tilde{r} = r\times \Bar{n}^{1/3}$ is the separation distance re-scaled by the cube root of the number density, $\Bar{n}^{1/3}$, so that $a, b$, $c$, $d$, and $e$ become dimensionless parameters.

\begin{figure*}
    \centering \includegraphics[width=0.95\linewidth]{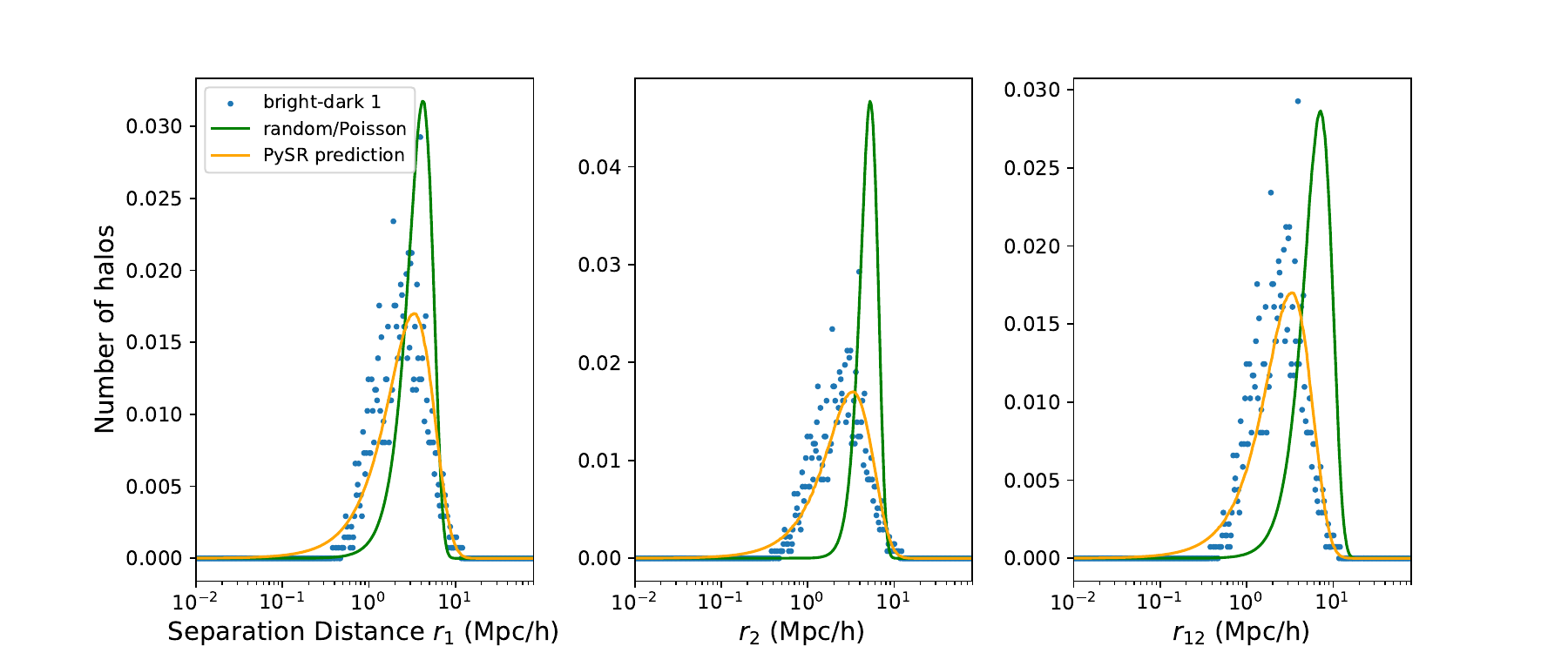}
    \caption{\footnotesize The probability density of predicting a halo in each distance bin depending on the separation distances of halos with their closest bright neighbour, second closest neighbour, and the two neighbours' distance to each other.  We can see that the ML prediction is more accurate than the analytical one for simulated halos rather than random sampling.}
\label{fig:sep_ml}
\end{figure*}

The $p(\Vec{r})$ found also fits the data well when multiplied by the probability density exclusion (Eq. \ref{eqn::theta}) - which comes from the triangle inequality.  This applies to galaxy-galaxy and galaxy-halo probability distributions as well - these probability functions can all take the same form but have different parameters (see Table \ref{tab::params}).  Figure \ref{fig:sep_ml} shows the result of Eq. \ref{ml_func} applied to the halo populations from the Illustris-2 simulation run.

\begin{figure*}
    \centering \includegraphics[width=0.95\linewidth]{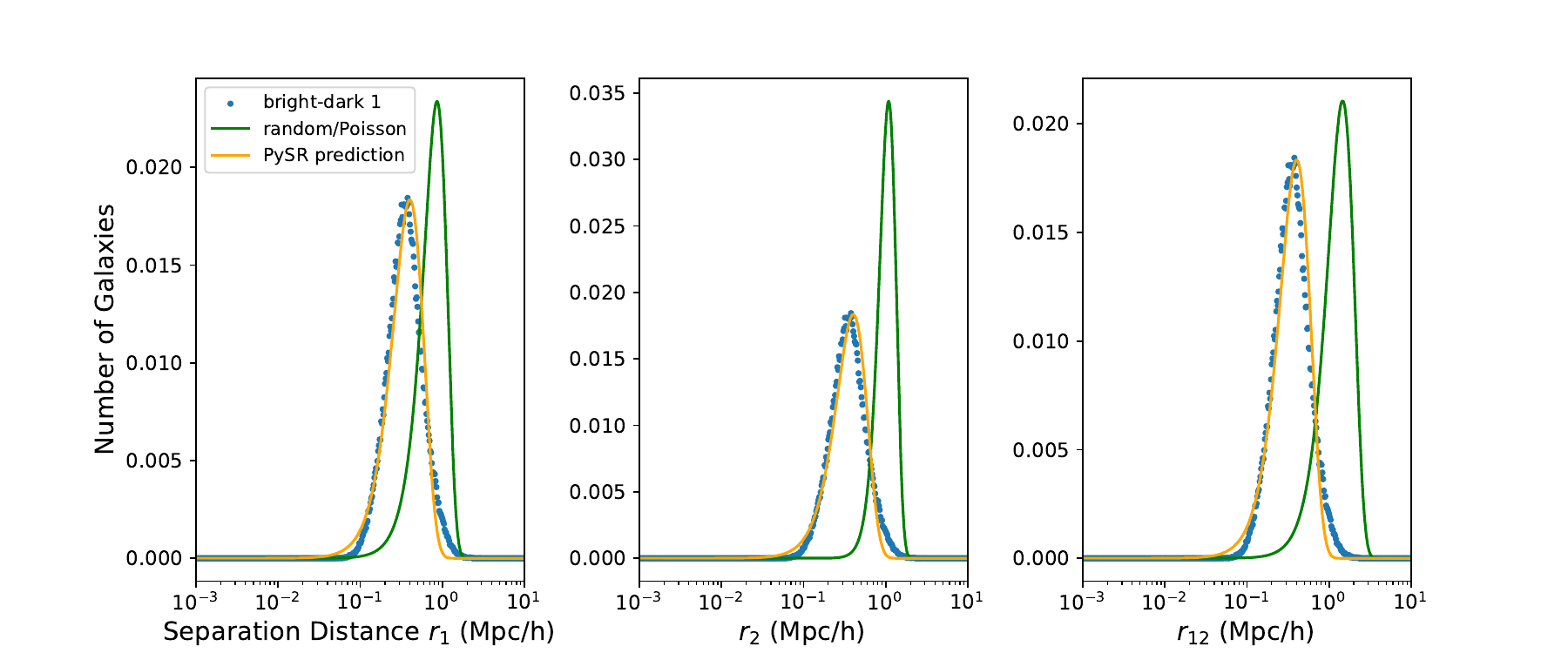}
    \caption{\footnotesize The probability density of finding a galaxy in each distance bin, based on separation distances of galaxies with their closest bright galactic neighbour, second closest neighbour, and the two closest neighbours with each other.  We can see that this data can be also fitted using symbolic regression, and if we tailor it to the data we obtain a better fit.}
\label{fig:sep_sub}
\end{figure*}

Because the halos and galaxies have different properties (mass, luminosity, pressure, etc.) that will affect their clustering, their probability density functions will have different parameters.  These parameters can be found using a numerical fitting method, such as Scipy's optimize package, or MCMC - both of which are faster than running the full symbolic regression algorithm again (reduces computational load).  The key point here is to show Eq. \ref{ml_func} can be used to model object distribution for both halos and subhalos, which implies the numerical expression has a certain degree of robustness.  If different parameters that provide good fits exist, then it should be possible to find them with any numerical algorithm, and that is what is shown in Table \ref{tab::params} below (here, we use NumPy's curve fit to find the parameters).  The results for galaxy-galaxy distribution probability are shown in Figure \ref{fig:sep_sub}.

\begin{figure*}
    \centering  \includegraphics[width=1.0\linewidth]{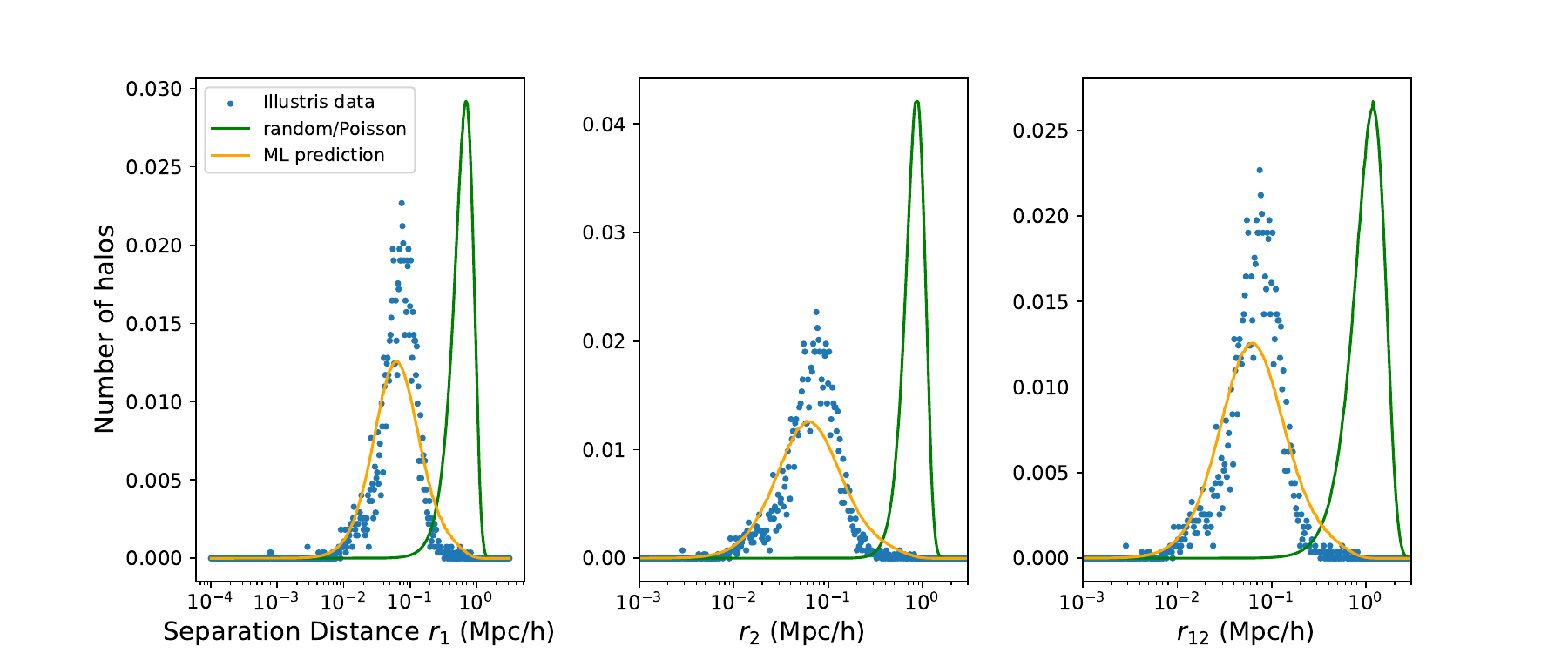}
    \caption{\footnotesize 
    The probability density of predicting a halo in each distance bin, based on separation to their closest neighbouring galaxies.}
\label{fig:sep_halo_galaxy}
\end{figure*}

\begin{figure*}
    \begin{tabular}{|ccc|}
    \hline  \\     \includegraphics[width=0.25\linewidth]{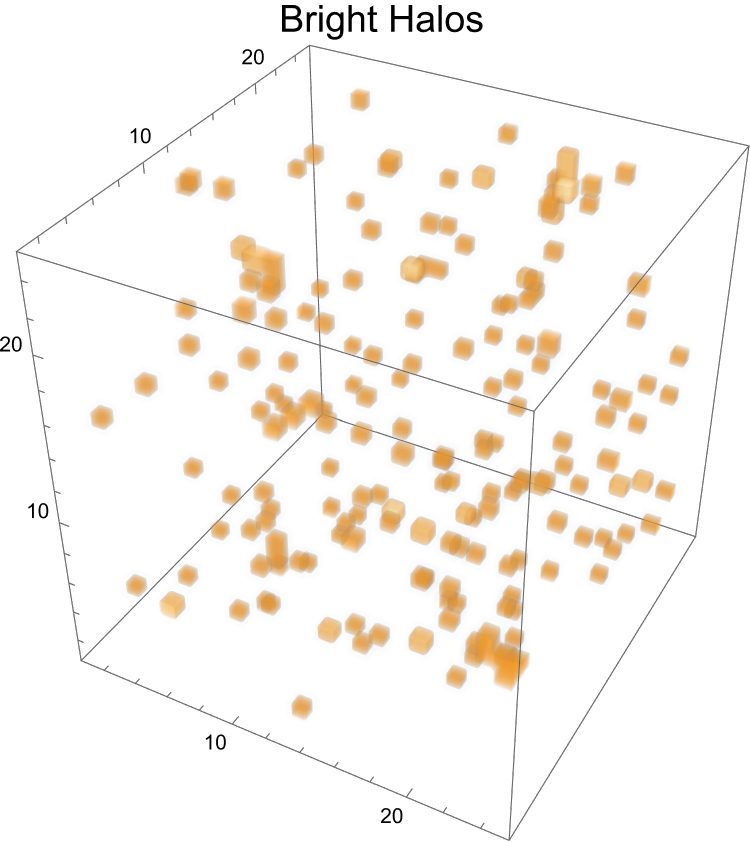}
     & \includegraphics[width=0.25\linewidth]{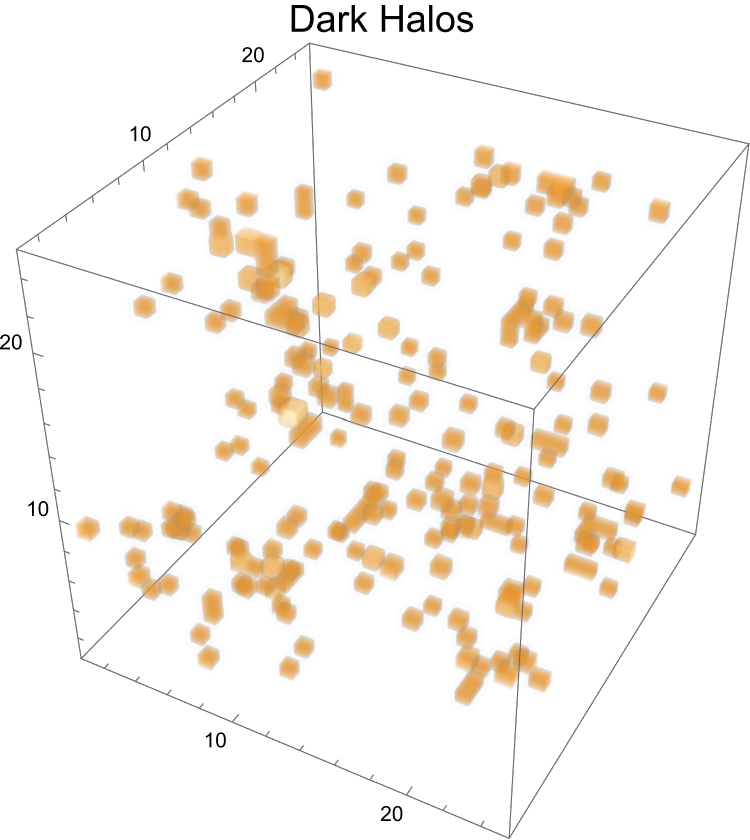} 
    & 
    \begin{subfigure}
    \centering \includegraphics[width=0.25\linewidth]{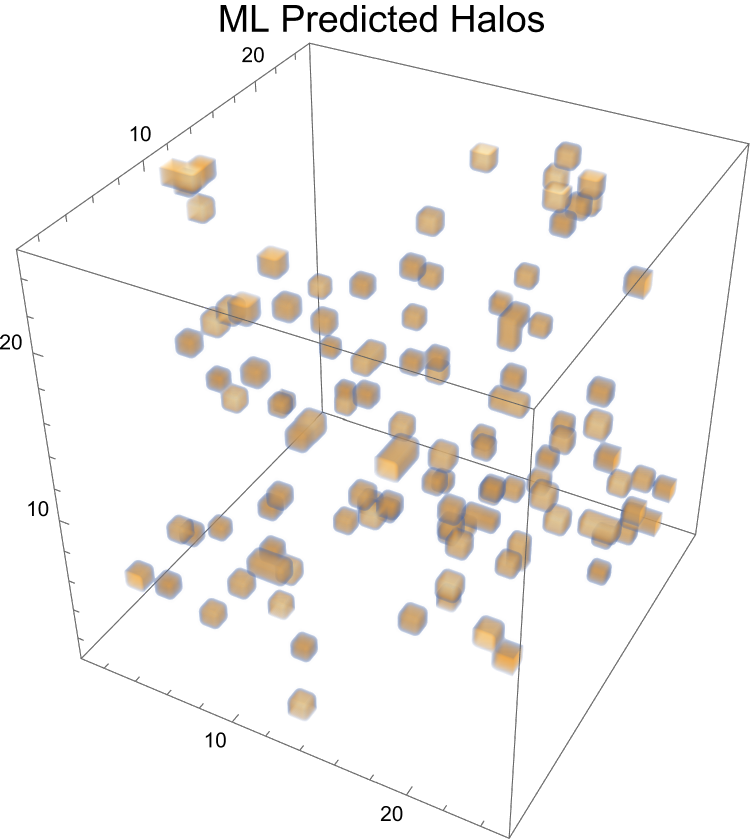}
    \end{subfigure}
    \begin{subfigure}
    \centering    \includegraphics[width=0.05\linewidth]{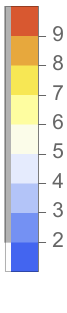}
    \end{subfigure}
    \\
    \hline    \includegraphics[width=0.25\linewidth]{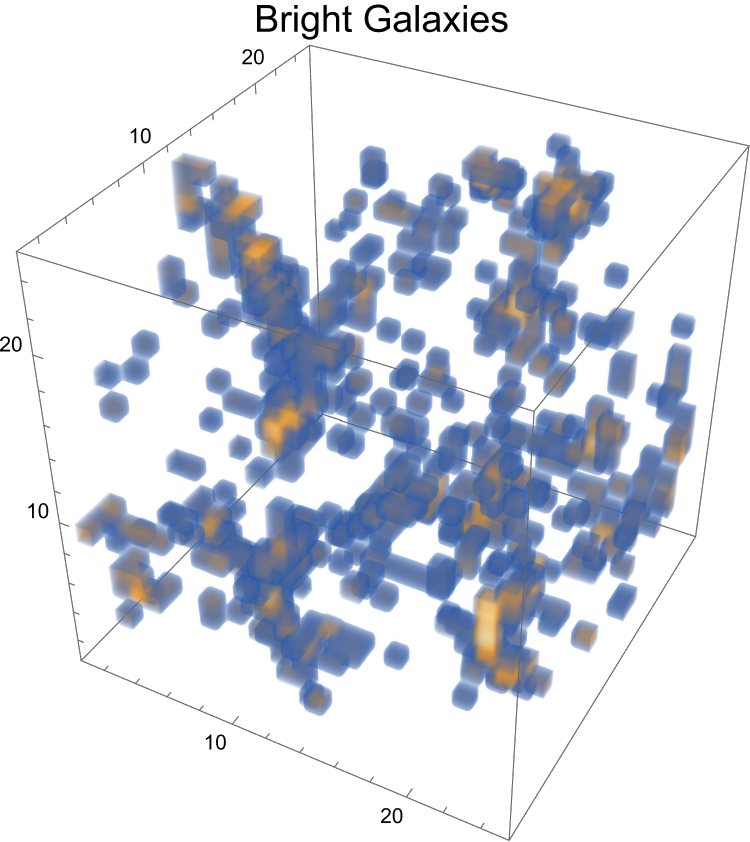}
    & \includegraphics[width=0.25\linewidth]{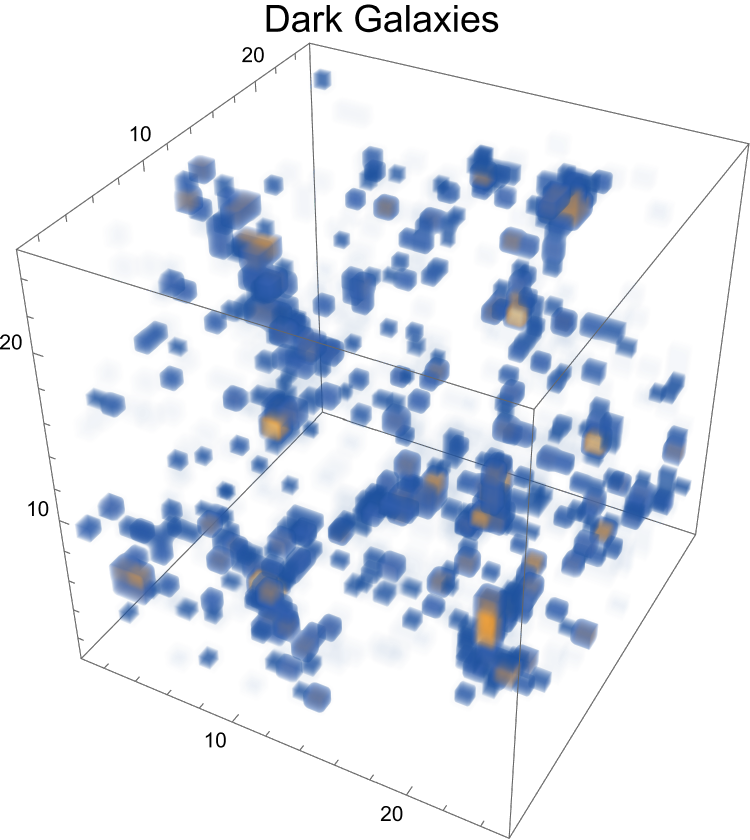} 
    & 
    \begin{subfigure}
    \centering \includegraphics[width=0.25\linewidth]{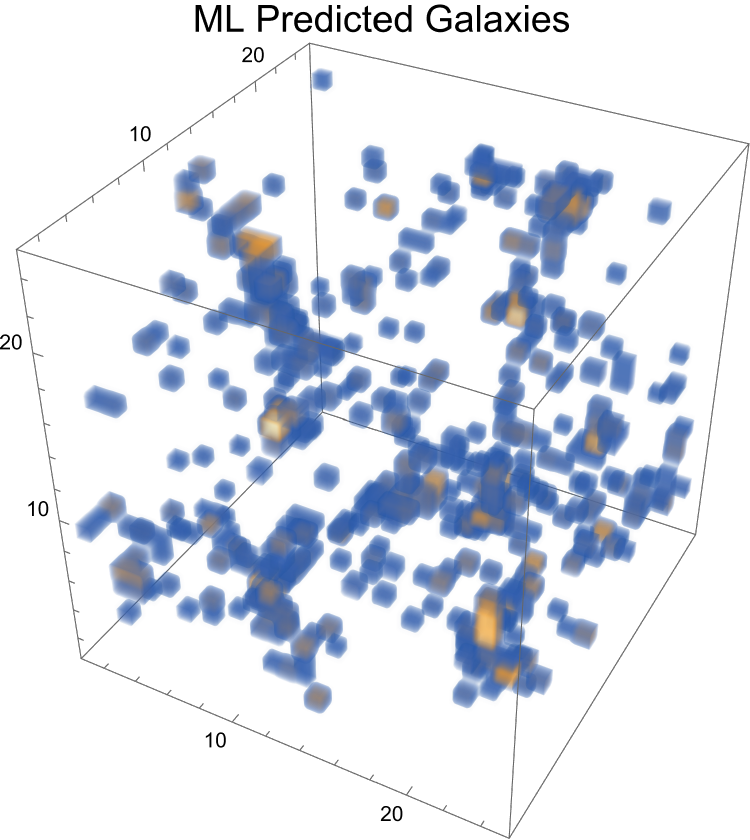}
    \end{subfigure}
    \begin{subfigure}
    \centering    \includegraphics[width=0.05\linewidth]{3d_colorbar_norm.pdf}
    \end{subfigure}
    \\
    \hline  \includegraphics[width=0.25\linewidth]{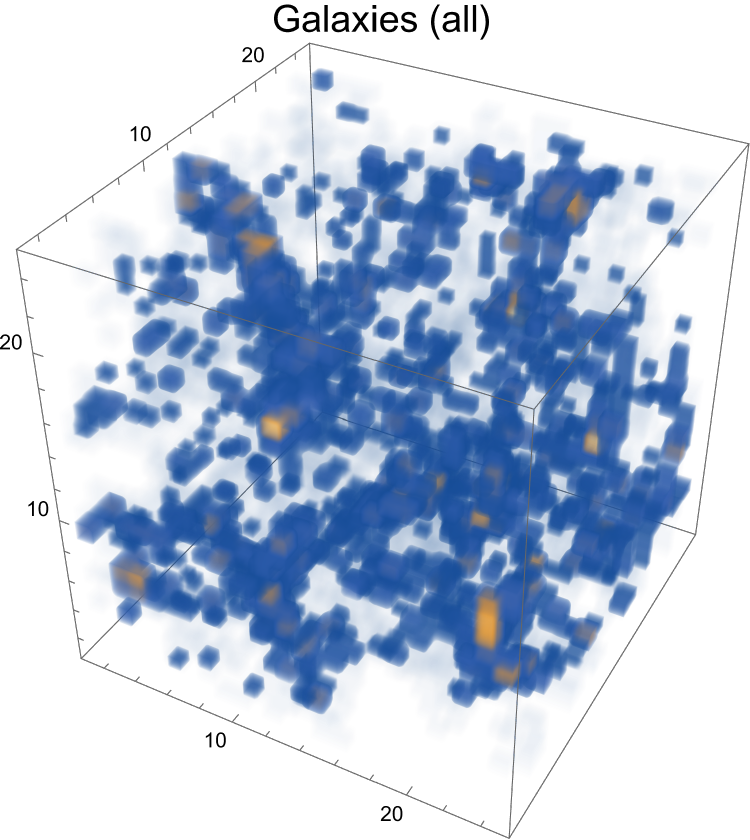}
    &    \includegraphics[width=0.25\linewidth]{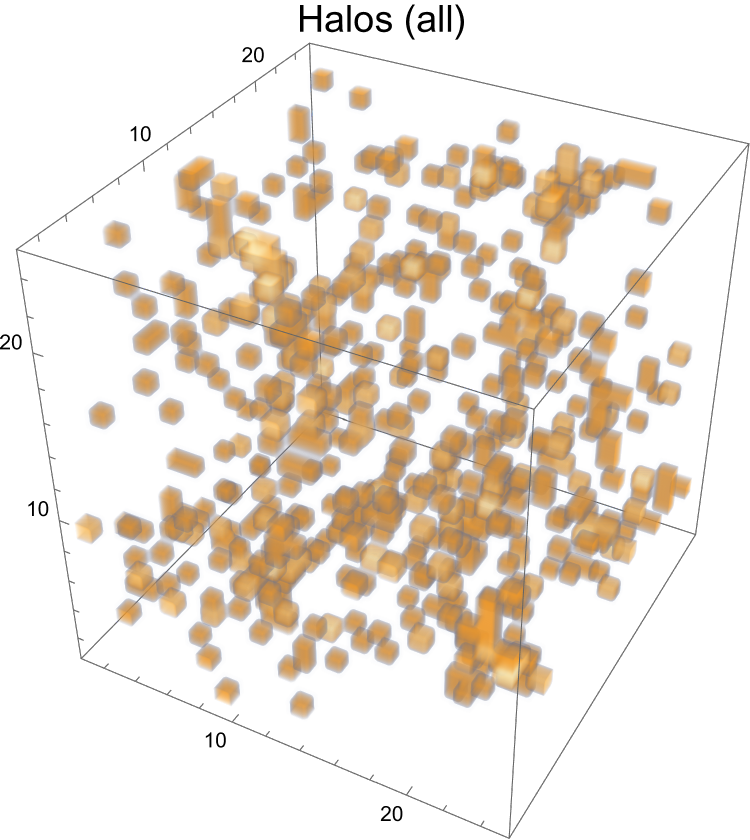} 
    &
    \begin{subfigure}
    \centering \includegraphics[width=0.25\linewidth]{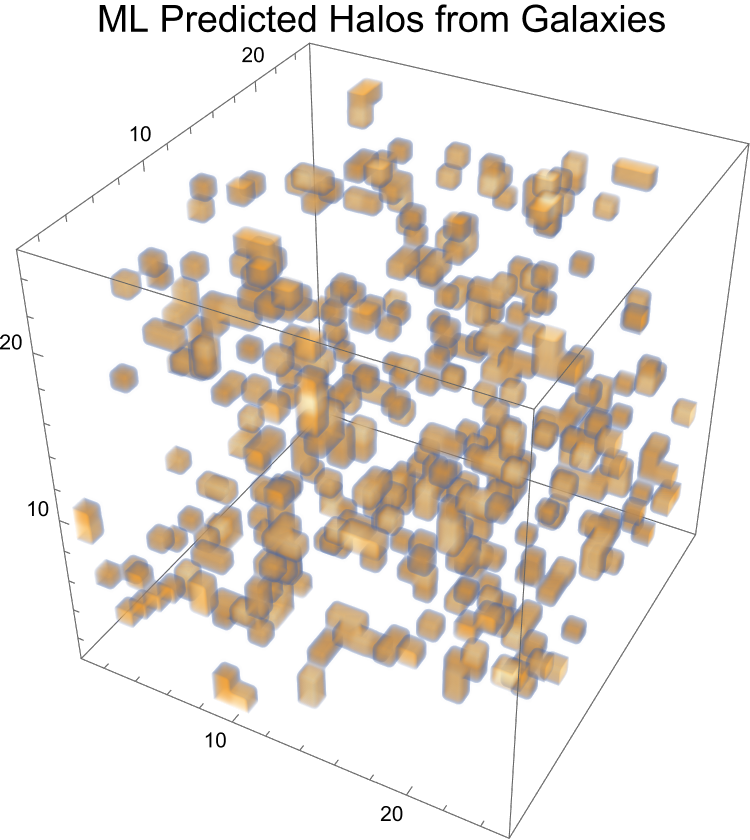}
    \end{subfigure}
    \begin{subfigure}
    \centering    \includegraphics[width=0.05\linewidth]{3d_colorbar_norm.pdf}
    \end{subfigure}
    \\
    \hline
    \end{tabular}
    \caption{\footnotesize  
    Comparison of the predicted 3D distribution of dark objects (depicted as number counts per 3 Mpc/h voxels) on the right column, with the their true distribution (middle column). The machine learning predictions are based on the distribution of bright objects, shown on the left column. On the top (middle) row, both bright and dark objects are dark matter haloes (galaxies), while in the bottom row, the distribution of dark haloes are predicted using bright galaxies. The axes are in units of voxel size.}
\label{fig:halo_3d}
\end{figure*}

Our final test sample is between galaxies, where the separation distance is now between the halo center and its closest two galaxies.  These ``closest" galaxies likely reside within the halos themselves, and their positions could make it easier to predict the halo centre (as well as potentially model substructure distribution in the future).  Here, we still use Eq. \ref{ml_func} for halo-halo and halo-galaxy probability, again with different parameters.  The results are summarized in Figure \ref{fig:sep_halo_galaxy}, with all the parameters listed in Table \ref{tab::params}.  
Figure \ref{fig:halo_3d} illustrates an example of predicting 3D densities of dark objects, derived from the distribution of bright objects using our symbolic regression approach, and compares this with the actual distribution of dark objects.

\begin{table*}
    \centering
    \begin{tabular}
     {|c|c|c|c|c|c|}
     \hline 
     Neighbouring Object & $a$ & $b$ & $c$ & $d$ & $e$ \\ 
     \hline
     Halo-Galaxy & 0.28 & 0.057 & 0.1 & 0.1 & 0.01 \\
     \hline
     Halo-Halo & 10.0 & 28.0 & 0.5 & 2.7 & 1.7 \\
     \hline 
     Galaxy-Galaxy & 180 & 210 & 1.0 & 8.0 & 0.7 \\
     \hline
    \end{tabular}
    \caption{\footnotesize The parameters for the fitting function Eq. \ref{ml_func} for each system: halo-galaxy, halo-halo, and galaxy-galaxy.} 
    \label{tab::params}
\end{table*}

\begin{figure*}
    \centering    \includegraphics[width=1.0\linewidth]{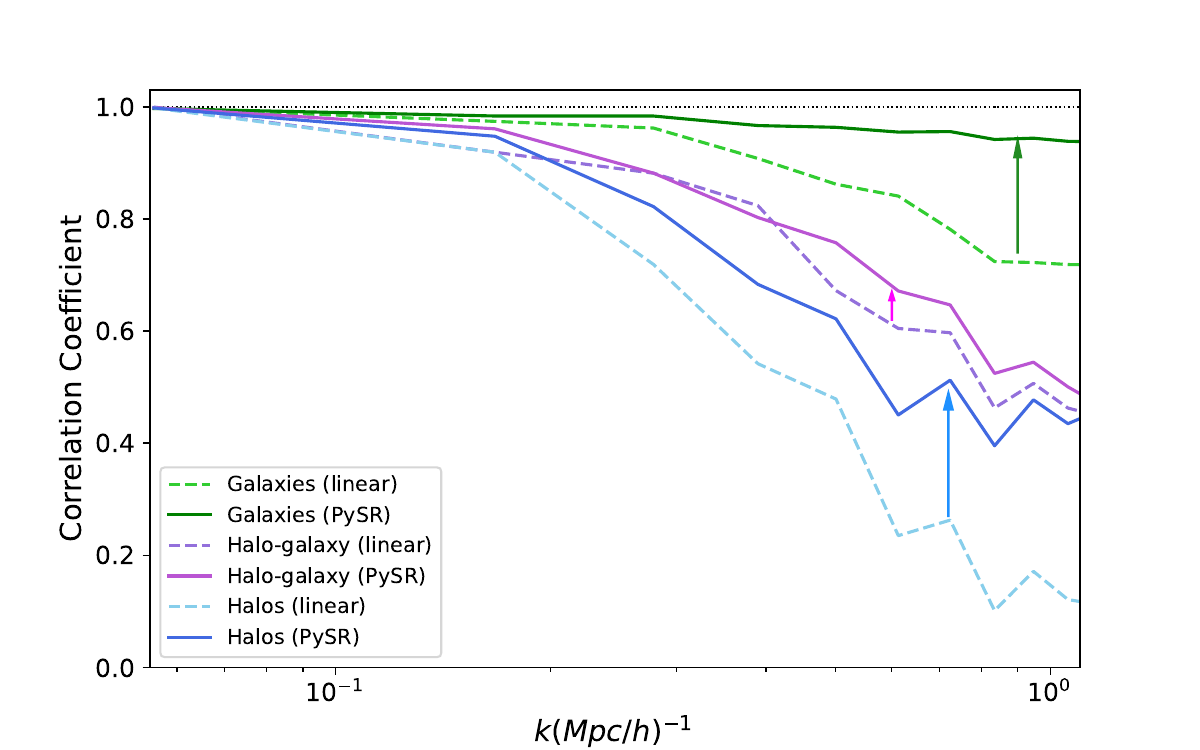}
    \caption{\footnotesize The correlation coefficients between the true 3D dark halo/galaxy map and our machine learning model based on 3D bright objects.  As can be seen from the graph, the machine learning model predicts the dark object distribution better than the map of bright/``visible" object raw number counts.  For the halo-galaxy correlation, we are using the galaxies as bright objects and halos are dark objects.}
    \label{fig:2PCF}
\end{figure*}

From our analysis, we see that the closest separation distances between halos and galaxies seem to peak around 1-3 Mpc, which consistent with the linking length found in \cite{Wu:2024jyu}.

\section{Discussion and Conclusion}
\label{sec:conclusion}
The results above show that it is possible to model the probability of finding halos and galaxies in the vicinity of each other as a function of their separation distances.  This probability function can be found using symbolic regression - shown in Eq. \ref{ml_func} - and it can be further verified with random forest algorithms and the loss/PIC (see Figures \ref{fig:halo_loss} and \ref{fig:rand_forest}), which shows that separation distances are significant when modelling neighbouring probability.  While the parameters of the probability function (see Table \ref{tab::params}) will change depending on what object (halo or galaxy) probability we are trying to predict, the form stays constant, showing that it is able to predict a considerable range of gravitationally clustered objects.  A 3D visualization of the dark object number count per voxel predicted by Eq. \ref{ml_func} is shown in the right column of Figure \ref{fig:halo_3d}; the column on the right is the predicted reconstruction of the middle column.  In all the graphs in Figure \ref{fig:halo_3d}, we applied a Heaviside function to remove residue for extremely small number density predictions ($\leq$1 object per voxel, since in reality there can't exist fractional galaxies/halos).

An immediate application of this approach is to help locate undetected dim galaxies in the vicinity of brighter galaxies (see Figures \ref{fig:halo_contours} and \ref{fig:sep_sub}).  This could be useful for finding galaxies that are hard to detect directly, but are theoretically predicted by the $\Lambda$CDM model  and important for studying structure formation \cite{Simon_2019} (e.g., the historically-debated ``missing satellites" problem or any remaining tensions despite advances in simulation resolution and data \cite{bullock2010notes, Sales:2022ich}, \cite{Homma:2023ppu}).  The same bright galaxies could also be used to infer the host halo center locations, which can further be used to predict where other halos would be found in their vicinities.  This could provide maps for where we should look for dark matter structure in weak lensing observations, in the context of a halo model for dark matter distribution. 

One extension of this galaxy/halo finding is to calculate the probability of finding a dark siren (with modest sky localization) in a constrained region of space, depending on which galaxies are contained within that region.  Given the observed galaxy distribution, it is possible to infer the probability distribution of finding a siren within that region, and this could be used to potentially localize dark sirens to a greater degree of accuracy (for applications such as measuring the Hubble parameter \cite{Finke:2021aom}).  

Another extension would be to compare our model to observational data.  While it is difficult to match simulation results to observations using a 1-1 mapping, comparing the distribution functions found from simulations to real data is easily feasible.  This could help further validate the accuracy of $\Lambda$CDM with regards to structure formation in observations.

Further progress would involve constraining the complexity and number of parameters for the model, as well as giving physical motivations as to why these objects would be distributed in the way suggested by the fitted probability function.  Moreover, we could try including luminosities of neighbouring galaxies in addition to position, and see if there is a relationship there that could be exploited to localize faint galaxies.  We should also point out that this work assumes a one-to-one baryon mass-luminosity dependence for galaxies and an unknown mass-luminosity dependence for halos where mass is the total mass - it would be interesting to see how the probability function in Eq. \ref{ml_func} would hold if we applied different mass definitions for ``bright" and ``dark" galaxies, for instance, stellar or gas mass instead of total mass.

This analysis can also be extrapolated to $k$NN statistics, where $k$ can be arbitrarily high instead of being constrained to $k=2$ as we have done here.  If we can accurately predict how many galaxies are in the vicinity of an observed bright galaxy using separation distance, then that's a signal that this model, as well as numerical simulation results, are good approximations to real space galaxy distributions.  This analysis would be a future follow-up work to this current paper and model.

\begin{acknowledgments}
We thank Miles Cranmer, David Spergel, Francisco Villaescusa-Navarro, Elizaveta Sazonova, and J. Luna Zagorac for helpful discussions and suggestions.

This research is supported by the Natural Sciences and Engineering Research Council of Canada, the University of Waterloo, and the Perimeter Institute for Theoretical Physics. Research at Perimeter Institute is supported in part by the Government of Canada through the Department of Innovation, Science and Economic Development and by the Province of Ontario through the Ministry of Colleges and Universities.
\end{acknowledgments}

\appendix

\section*{Appendix}
\label{sec:appendix}

\begin{figure}
   \centering \includegraphics[width=0.95\linewidth]{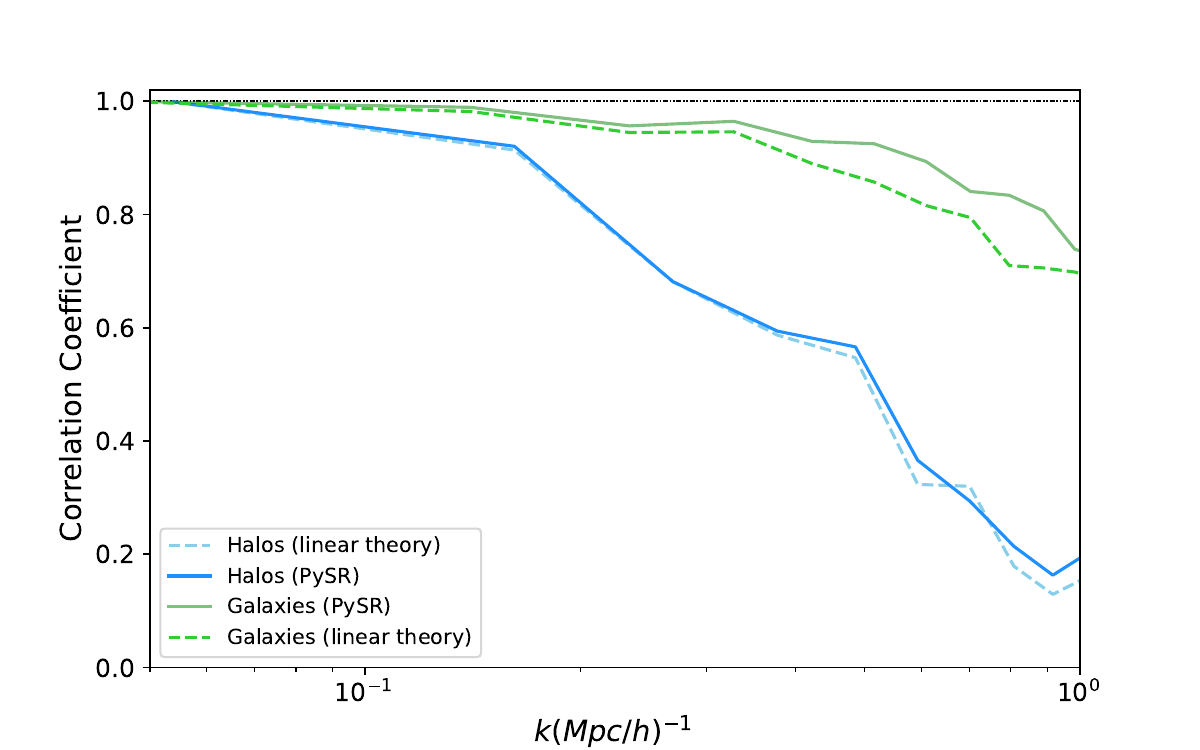}
   \caption{\footnotesize The model from Eq. \ref{ml_func} with TNG-100 data.  Here, we only have halo-halo and galaxy-galaxy since TNG-100 has slightly less halos than Illustris-2 did, but much more galaxies, so halo voxel size is different from galaxy voxel size, and thus a halo-galaxy comparison isn't done in this case.  We can see that while some parameters now have uncertainties to them, the fitting function itself can still provide a fit than the linear theory, especially for galaxies.}
\label{fig:tng_cf}
\end{figure}

\begin{figure}
   \centering \includegraphics[width=0.92\linewidth]{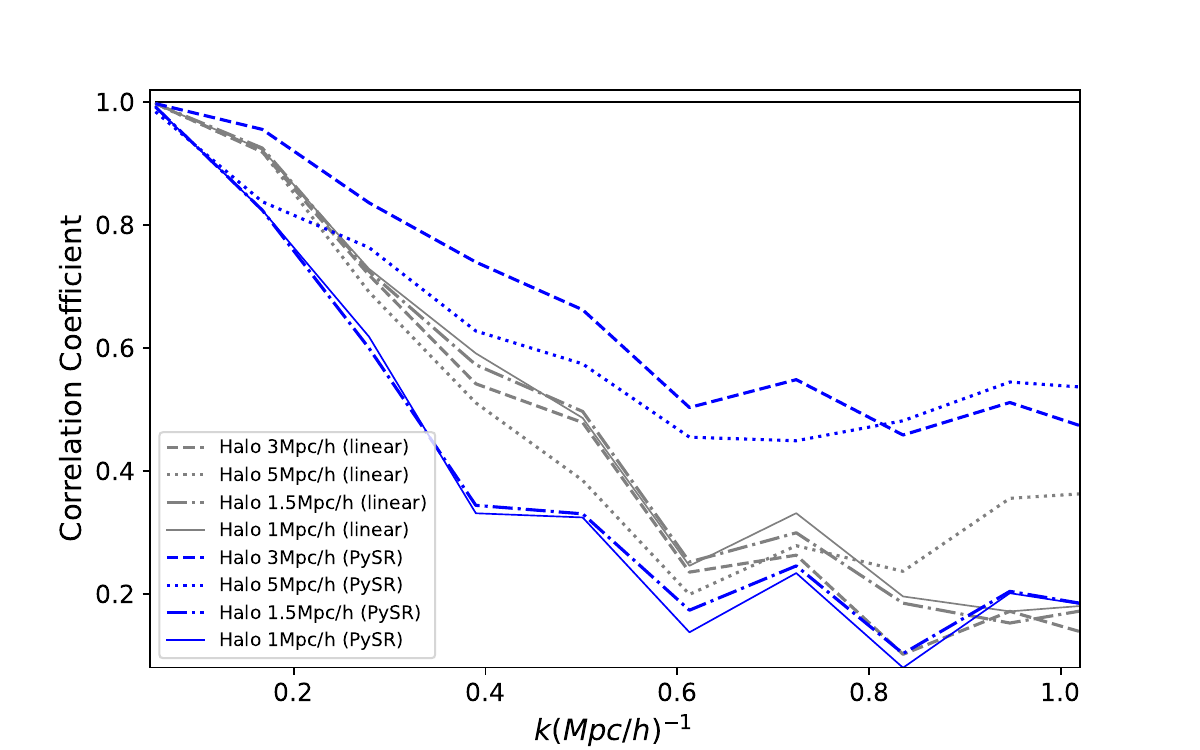}
   \caption{\footnotesize The difference voxel size, and thus number density, makes in the correlation functions, since object number density remains constant.  We can see that generally $\geq$1 object per voxel fits the data best.}
\label{fig:all_voxel_cf}
\end{figure}

Here, we show further examples of the PySR model Eq. \ref{ml_func} with different sets of simulation data.  The data shown below is from the TNG-100 simulation \cite{Nelson:2018uso}, with slightly different cosmological parameters than Illustris-2 (Planck 2015 vs WMAP-9).  While this gives us some uncertainty on the parameters, we can still see that the symbolic regression model gives us a better fit than the linear model for neighbouring galaxies, shown in Figure \ref{fig:tng_cf}.  This also reaffirms the trend we see in Figure \ref{fig:all_voxel_cf}, where number density per voxel has to be within a certain range for a good numerical fit, approximately 0.9-5 objects per voxel.  In both Illustris-2 ($\sim$3300 halos, $\sim$15000 galaxies) and TNG-100 ($\sim$3100 halos, $\sim$75000 galaxies) results, the galaxy model has a better fit compared to halos.  This is likely due to there being a higher object number density compared to halos, shown in Figure \ref{fig:all_voxel_cf} where we have the same number of objects from Illustris-2, but different voxel sizes.  This shows that the model found by PySR in Eq. \ref{ml_func} has a certain robustness to it; even if the parameter needs changing, the general form of this equation can be used to model objects exhibiting gravitational clustering.

\begin{table}
    \centering
    \begin{tabular}
     {|c|c|c|c|c|c|}
     \hline 
     Neighbouring Object & $a$ & $b$ & $c$ & $d$ & $e$  \\ 
     \hline
     Halo-Halo & 10.0 & 28.0 & 0.01 & 1.2 & 0.1 \\
     \hline 
     Halos Average & 10.0 & 28.0 & 0.26 & 1.9 & 0.9 \\
     \hline 
     Halos ($\sigma$) & 0.0 & 0.0 & 0.24 & 0.7 & 0.8 \\
     \hline 
     Galaxy-Galaxy & 180 & 210 & 0.02 & 1.2 & 0.2 \\
     \hline
     Galaxies Average & 180 & 210 & 0.51 & 4.6 & 0.45 \\
     \hline 
     Galaxies ($\sigma$) & 0.0 & 0.0 & 0.5 & 3.4 & 0.25 \\
     \hline 
    \end{tabular}
    \caption{\footnotesize The parameters, including averages and standard deviations for the fitting function Eq. \ref{ml_func} for halo-halo and galaxy-galaxy distributions based on both the Illustris-2 and TNG-100 results.} 
    \label{tab::params_2}
\end{table}

Furthermore, we shown in Table \ref{tab:all_eqns} below that the equations all converge to a form very similar to Eq. \ref{ml_func} during PySR's modelling process, showing that gravitationally clustered objects tend to follow the fraction $\times$ inverse exponential trend we found in our model above.

\begin{figure}
   \centering \includegraphics[width=0.92\linewidth]{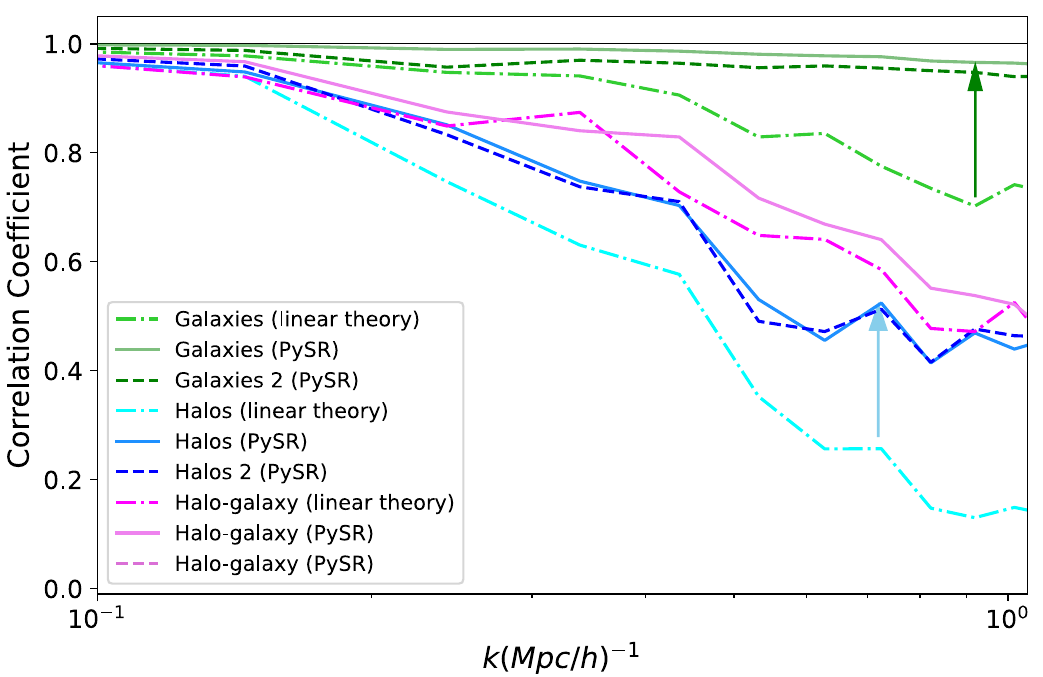}
   \caption{\footnotesize The model from Eq. 12 in Table \ref{tab:all_eqns}, since it has a PIC minimum.  However, as it is missing the closest neighboring distance (and from gravitational theory, likely the most important distance), does not have a viable fit for halo-galaxy modelling, and does worse in galaxy-galaxy modelling (though comparable to our current model in halo-halo modelling), we still use Eq. \ref{ml_func} in the rest of this work.}
\label{fig:model2_cf}
\end{figure}

\begin{table*}
    \begin{tabular}{|p{0.05\linewidth}| p{0.15\linewidth}|p{0.7\linewidth}|}    
    \hline  \\
    Eq. & Complexity & $\ln[Equation]$ \\ 
     \hline
    0 & 1 & 2.41 \\ 
     \hline
    1 & 3 & $\ln{\Big| \cfrac{1}{x_{1}} \Big|}$ \\ 
     \hline
    2 & 5 & $0.860\ln{\Big| \cfrac{1}{x_{1}} \Big|}$ \\ 
     \hline
    3 & 7 & $(x_{1} + 0.860) \ln{\Big| \cfrac{1}{x_{1}} \Big|}$ \\ 
     \hline
    4 & 9 & $\ln{\Big|\cfrac{1}{(x_{0} + 0.734) (x_{1} + x_{2})} \Big|}$ \\ 
     \hline
    5 & 11 & $\ln{\Big|\cfrac{1}{(x_{1} + x_{2}) (x_{1}^{2} + 0.743)} \Big|}$ \\ 
     \hline
    6 & 13 & $(x_{1} + 0.930) \ln{\Big|\cfrac{1}{2 x_{1} + x_{2}} \Big|} + 0.849$ \\ 
     \hline
    7 & 15 & $(x_{1}^{2} + 0.912) \ln{\Big|\cfrac{1}{2 x_{1} + x_{2}} \Big|} + 0.960$ \\ 
     \hline
    8 & 16 & $(x_{0} + 0.894) \ln{\Big|\cfrac{1}{2 x_{1} + x_{2}} \Big|} + \cfrac{1}{x_{1} + 0.981}$ \\ 
     \hline
    9 & 17 & $(x_{1}^{2} + x_{2} + 0.917) \ln{\Big|\cfrac{1}{2 x_{1} + x_{2}}\Big|} + 0.900$ \\ 
     \hline
    10 & 18 & $x_{2} + (x_{1} + 0.919) \ln{\Big|\cfrac{1}{2 x_{1} + x_{2}} \Big|} + \cfrac{1}{x_{1} + 1.12}$ \\ 
     \hline
    11 & 20 & $(x_{1}^{2} + 0.891) \Big(x_{2} + \ln{\Big|\cfrac{1}{2 x_{1} + x_{2}} \Big|)} \Big) + \cfrac{1}{x_{1} + 0.954}$ \\ 
     \hline
    12 & 22 & $(x_{1}^{3} + 0.887) \Big(x_{2} + \ln{\Big|\cfrac{1}{2 x_{1} + x_{2}} \Big|} \Big) + \cfrac{1}{x_{1} + 0.940}$ \\ 
     \hline
    13 & 24 & $(x_{1}^{4} + 0.887) \Big( x_{2} + \ln{\Big|\cfrac{1}{2 x_{1} + x_{2}} \Big|} \Big) + \cfrac{1}{x_{1} + 0.939}$ \\ 
     \hline
    14 & 30 & $\Big( x_{2} + \ln{\Big|\cfrac{1}{2 x_{1} + x_{2}} \Big|} \Big) (x_{1} \cdot (1.88 x_{1} (x_{1} - 0.612) + 0.263) + 0.894) + \cfrac{1}{x_{1} + 0.971}$ \\ 
     \hline
    15 & 32 & $x_{2} + (x_{0} x_{1}^{2} (x_{1} - 1.37 x_{2}) + 0.884) \ln{\Big|\cfrac{1}{2 x_{1} + x_{2}} \Big|} + \cfrac{1}{x_{1}^{2} + x_{1} + 0.926}$ \\
    \hline
    \end{tabular}
    \caption{\footnotesize  The list of equations from PySR for the simulation data we are using.  In our notation, $x_0 \rightarrow \Tilde{r}_1 = \Bar{n}^{1/3}_b r_1$, $x_1 \rightarrow \Tilde{r}_2 = \Bar{n}^{1/3}_b r_2$, and $x_2 \rightarrow \Tilde{r}_{12}= \Bar{n}^{1/3}_b r_{12}$.  The equation we ended up using was on line 8, picked by using the methods shown in Figure \ref{fig:halo_loss}.}
    \label{tab:all_eqns}
\end{table*}

\clearpage

\bibliographystyle{plain}
\bibliography{references}

\end{document}